\documentclass[prc,twocolumn,aps,showpacs,preprintnumbers, tightenlines,showpacs,floatfix,superscriptaddress,nofootinbib]{revtex4-1}

\usepackage{bm}
\usepackage{color}
\usepackage{hyperref}
\usepackage{graphicx,epsfig}
\usepackage{longtable}
\usepackage{xcolor}
\usepackage{color}
\usepackage{float}

\usepackage{dcolumn}     %
\usepackage[resetlabels,labeled]{multibib}
\newcites{Math}{Mathgg}

\def\beq{\begin{equation}}
\def\eeq{\end{equation}}
\def\beqy{\begin{eqnarray}}
\def\eeqy{\end{eqnarray}}

\newcommand{\pkuphy}{School of Physics, Peking University, Beijing 100871,
China}
\newcommand{\gscaep}{Graduate School of China Academy of Engineering Physics, Beijing 100193, China}
\newcommand{\chep}{Center for High Energy Physics, Peking University, Beijing 100871, China}
\newcommand{\ccqm}{Collaborative Innovation Center of Quantum Matter, Beijing 100871, China}

\begin{document}

{
%\title{Investigating Nuclear Beta Decay Using Nuclear Lattice Effective Field Theory}

\title{Investigating nuclear $\beta$ decay using a lattice quantum Monte Carlo approach}

%\title{}

\author{Teng~Wang}\email{tenggeer@pku.edu.cn}\affiliation{\pkuphy}
\author{Xu~Feng}\email{xu.feng@pku.edu.cn}\affiliation{\pkuphy}\affiliation{\chep}\affiliation{\ccqm}
\author{Bing-Nan~Lu}\email{bnlv@gscaep.ac.cn}\affiliation{\gscaep}

\begin{abstract}
%We present a numerical study of nuclear $\beta$-decay using the Monte Carlo approach within the framework of Nuclear Lattice Effective Field Theory, focusing on the $\beta$-decays of $^3$H and $^6$He. 
%Nuclear forces, along with one- and two-body axial currents consistently derived from chiral effective field theory, are employed to calculate the reduced Gamow-Teller matrix element.
%First, we determine the low-energy constants $c_D$ and $c_E$ in the three-body nuclear force by combining the half-life of triton $\beta$-decay with the triton binding energy. 
%Using these calibrated constants, we then predict the matrix element for $^6$He $\beta$-decay, finding reasonable agreement with experimental data.

%We present 
We present an \textit{ab initio} calculation of nuclear $\beta$ decay within the framework of nuclear lattice effective field theory (NLEFT), employing auxiliary-field quantum Monte Carlo methods to solve the nuclear many-body problem. Our approach combines next-to-next-to-leading order  two- and three-body chiral interactions with one- and two-body axial current operators, all consistently derived in chiral effective field theory. Low-energy constants are determined exclusively from nucleon-nucleon scattering phase shifts and few-body observables for systems with $A \leq 3$. Using these interactions and transition operators, we perform two-channel Monte Carlo simulations to compute the $\beta$-decay matrix element for $^6$He, obtaining results in reasonable agreement with experimental measurements. To address the Monte Carlo sign problem, we implement a perturbative expansion around a leading-order Hamiltonian with approximate Wigner-SU(4) symmetry. This systematic approach provides a foundation for extending NLEFT simulations to precision studies of weak processes in medium-mass nuclei.

\end{abstract}

\maketitle
}

\section {Introduction}
\label{sec:intro}

Nuclear $\beta$ decay is a fundamental process crucial to understanding the laws of nature. Historically, studies of $\beta$ decay have led to 
groundbreaking discoveries, including the existence of neutrinos~\cite{Pauli:1978}, the confirmation of parity violation~\cite{Wu:1957}, and contributions to the unification of electroweak interactions~\cite{Glashow:1961,Weinberg:1967,Salam:1968}.
Today, nuclear $\beta$ decay provides a sensitive probe for physics beyond the standard model. 
For a comprehensive review of ongoing and planned experimental programs, see Ref.~\cite{Alonso:2019} and references therein.

Besides its central role in particle physics, nuclear $\beta$ decay is also of great interest within the field of low-energy nuclear physics. 
Currently there remain several unresolved challenges in precisely describing the nuclear weak decay with the nucleonic degrees of freedom.
For instance, traditional shell-model calculations tend to overestimate the 
reduced Gamow-Teller matrix element (RGTME), especially for medium-mass nuclei~\cite{Brown:1985,Pinedo:1996,Langanke:1995}.
To reconcile this discrepancy, the axial charge $g_A$ in the axial current is often tuned to match experimental data~\cite{Chou:1993}, leading to the well-known $g_A$ quenching problem.
Moreover, the nuclear force binds nucleons into atomic nuclei, inducing complicated many-body correlations, which requires tremendous amount of computing resources and advanced algorithms to solve the corresponding quantum many-body problems.   
Lastly, the external electroweak field interacts with the nucleus via not only  one-body currents but also many-body currents, so a proper description of $\beta$ decays requires a consistent treatment of the electroweak currents together with the  complicated nuclear chiral interactions.

Recent years have seen rapid development of $ab \ initio$ nuclear many-body methods~\cite{QMCreview,NCSMreview,CCreview,IMSRGreview,Lee:2009}, which provide a consistent and systematically improvable framework for understanding the intricate interplays among nuclear forces, electroweak currents, and quantum many-body correlations in $\beta$-decay processes. 
For nuclei of $s$ and $p$ shells, calculations based on the quantum Monte Carlo (QMC) method~\cite{Pastore:2018,King:2020} and the no-core shell model method~\cite{NCSM_A10_13,Gysbers:2019} have been performed systematically,  with the latter applied to the study of the anomalous long  lifetime of $^{14}$C~\cite{NCSM_C14}. For heavier nuclei, the coupled-cluster method  has been applied to investigating the impact of three-body forces and two-body currents on the Gamow-Teller transitions of $^{14}$C, $^{22}$O and $^{24}$O~\cite{CC_C14_O22_O24}, and it also provided the first $ab$ $inito$ calculation of  $^{100}$Sn $\beta$ decay~\cite{Gysbers:2019}. Besides,  the valence-space in-medium similarity renormalization
group method has made significant progress in studying  $\beta$ decays of medium-mass nuclei~\cite{Gysbers:2019,VSIMSRG_2021,VSIMSRG_Co53}.
%as well as $^{100}$Sn recently~\cite{VSIMSRG_Co53}.   
These $ab$ $initio$ calculations have significantly improved our understanding of nuclear $\beta$ decays, yet also raised a lot of open questions. 
For instance, the QMC calculations struggle to reproduce the experimental $\beta$ decay rate for $^8$Li, $^8$B, and $^8$He~\cite{King:2020}. 
Furthermore, the contribution of two-body currents is found to have opposite signs in Refs.~\cite{King:2020} and~\cite{Gysbers:2019}. 
This theoretical inconsistency was further explored in Ref.~\cite{Gnech:2021}, which also revealed that the one-pion-exchange (OPE) current is highly sensitive to the nuclear forces and regularization schemes. 

As one of the advanced nuclear \textit{ab initio} methods, the nuclear lattice effective field theory (NLEFT)~\cite{Lee:2009,Lahde:2019} implements the chiral interactions on a cubic spatial lattice and solves the corresponding discrete many-body Schr\"odinger equation using the auxiliary-field Monte Carlo method.
Recently the NLEFT has achieved remarkable progress in studying the properties of nuclear ground states and excited
states~\cite{EPJA31-105,PhysRevLett.104.142501,Epelbaum:2010,PLB732-110,Lu:2019,Lu:2021,Elhatisari:2022, PhysRevLett.112.102501,Shen:2023,Ulf:2024,arxiV2411.14935,arxiV2411.17462,Teng:2025},
nuclear clustering~\cite{Epelbaum:2009a,PhysRevLett.109.252501,PhysRevLett.110.112502,Elhatisari:2016,Elhatisari:2017}, scattering processes~\cite{PhysRevC.86.034003,Elhatisari:2015}, thermal nuclear
matter~\cite{Lu:2020,Ren:2023,Ma:2023}
, and hypernuclei~\cite{PhysRevLett.115.185301,EPJA56-24,EPJA60-215}. In this area, in recent years a significant amount of effort was made to build high-precision lattice interactions and design algorithms that mitigate the Monte Carlo sign problem.
For example, the wave function matching method~\cite{Elhatisari:2022} and perturbative quantum Monte Carlo method~\cite{Lu:2021, arXiv2502.13565} were proposed to improve the accuracy of perturbative calculations around a simple Hamiltonian without the sign problem. 
When combined with high-fidelity nuclear forces, these methods were able to reproduce the experimental binding energies and charge radii with systematically controllable statistical errors.

Attempts to study nuclear $\beta$ decay through NLEFT have just begun very recently.
The first and currently the only study focuses on the triton $\beta$ decay~\cite{Elhatisari:2024gg}, which employs the recently developed high-fidelity next-to-next-to-next-to-leading order (N$^3$LO) chiral interaction~\cite{Elhatisari:2022} and the leading-order (LO) axial current to achieve consistent predictions for the Fermi and Gamow-Teller matrix elements. The sensitivity of these matrix elements to 
the strengths of three-body forces is also carefully examined therein.

In this work, we extend the work of~\cite{Elhatisari:2024gg}  by presenting an in-depth NLEFT study of nuclear $\beta$ decay, with major developments of two aspects. First, we have incorporated one- and two-body axial currents beyond LO in the calculation, which are essential for the determination of three-body low-energy constants (LECs), $c_D$ and $c_E$, of the next-to-next-to-leading order (N$^2$LO) chiral nuclear force through triton $\beta$ decay ~\cite{Gazit:2009}, and it has been shown that the two-body currents provide sizable contributions to  Gamow-Teller matrix elements of certain nuclei~\cite{King:2020, Gysbers:2019}.  Second, we have succeeded in simulating nuclear $\beta$ decay through the lattice quantum Monte Carlo method, which can be directly applied to weak decays of medium-mass and heavy nuclei. As a first step, here we focus on the two simplest $\beta$ decay processes, i.e., $^3$H\ $\rightarrow{^3}$He and $^6$He\ $\rightarrow{^6}$Li. We first fix $c_D$ and $c_E$ through the former process, then validate the applicability of our method using the latter. Since the main goal of this work is to build the relevant Monte Carlo
machinery for investigating nuclear $\beta$ decay through NLEFT, we use the N$^2$LO lattice chiral interaction fitted to nucleon-nucleon scattering phase shifts and few-body observables with $A\le 3$ for numerical convenience. A systematic study based on the high-fidelity N$^3$LO interaction is left for the future.  

This paper is organized as follows. The nuclear interaction and  axial 
currents used in this work are described in Secs.~\ref{sec:lat interaction} and~\ref{sec:current}. Our calculation methods and techniques are briefly discussed in Sec.~\ref{sec:cal method}. 
Results for $^3$H and $^6$He $\beta$ decay  are presented in Sec.~\ref{sec:Res}. More details are provided in the Appendixes.

\section{Interaction}
\label{sec:lat interaction}

\subsection{Full interaction}

The full Hamiltonian used in this work is the N$^2$LO chiral Hamiltonian discretized on a lattice with a lattice spacing $a=1.64$ fm,
\begin{equation}
    H = K+ V_{\mathrm{OPE}}^{\Lambda_\pi}+V_{C_\pi}^{\Lambda_\pi}+V_{2\mathrm{N}}^{Q^2}+V_{\mathrm{IB}}+V^{Q^3}_{3\mathrm{N}}.
    \label{eq:fullH}
\end{equation}
Here, $K$ is kinetic term, and the nucleon mass is taken as $M_N = 938.92$ MeV. The OPE potential is locally regularized in momentum space, following the prescription of Ref.~\cite{Reinert:2018},
\begin{equation}   V_{\mathrm{OPE}}^{\Lambda_\pi} = -\frac{\tilde{g}_A^2}{4f_\pi^2}(\boldsymbol{\tau}_1\cdot\boldsymbol{\tau}_2)\frac{\boldsymbol{(\sigma}_1\cdot\boldsymbol{q})(\boldsymbol{\sigma}_2\cdot\boldsymbol{q})}{\boldsymbol{q}^2+M_\pi^2}e^{-\frac{\boldsymbol{q}^2+M_\pi^2}{\Lambda_\pi^2}}, 
\label{2NOPE}
\end{equation}
\begin{equation}
V_{C_\pi}^{\Lambda_\pi} =  -\frac{\tilde{g}_A^2C_\pi}{4f_\pi^2}(\boldsymbol{\sigma}_1\cdot\boldsymbol{\sigma}_2)(\boldsymbol{\tau}_1\cdot\boldsymbol{\tau}_2),
    \label{eq:counterterm}
\end{equation}
where
\begin{equation}
    C_\pi = \frac{\Lambda_\pi(\Lambda_\pi^2-2M_\pi^2)+2\sqrt{\pi}M_\pi^3\exp\left(\frac{M_\pi^2}{\Lambda_\pi^2}\right)\mathrm{erfc}\left(\frac{M_\pi}{\Lambda_\pi}\right)}{3\Lambda_\pi^3}.
\end{equation}
The numerical values of the parameters used are $\tilde{g}_A = 1.287$ for the nucleon axial charge, $f_\pi = 92.2$ MeV for the pion decay constant, $M_\pi=134.98$ MeV for the  pion mass, and
$\Lambda_\pi = 300$ MeV for the local regulator.
The axial charge $\tilde{g}_A$ incorporates the Goldberger-Treiman discrepancy~\cite{Arndt:1994} and differs slightly from value $g_A = 1.2723(23)$~\cite{Tanabashi:2018} in the axial current.
To distinguish between them, we use a tilde notation.
The term in Eq.~(\ref{eq:counterterm}) serves as a counterterm to remove the short-distance singularity in the OPE potential. 
It is worth noting that the characteristic momentum scale in this work is around 100 MeV, considerably lower than twice the pion mass. 
As a result, the two-pion exchange (TPE) potential, which first appears at next-to leading order (NLO), cannot be distinguished from contact terms~\cite{Epelbaum_2009}.

The two-nucleon (2N) short-range potential, $V^{Q^2}_{2\mathrm{N}}$, includes two contact terms at leading order (LO) and seven additional contact terms at next-to-leading order (NLO),
\begin{eqnarray}   V_{2\mathrm{N}}^{Q^2}&=&B_1+B_2(\boldsymbol{\sigma}_1\cdot\boldsymbol{\sigma}_2)+C_1q^2+C_2 q^2(\boldsymbol{\tau}_1\cdot\boldsymbol{\tau}_2)\nonumber\\
&+&C_3 q^2(\boldsymbol{\sigma}_1\cdot\boldsymbol{\sigma}_2)+C_4 q^2(\boldsymbol{\sigma}_1\cdot\boldsymbol{\sigma}_2)(\boldsymbol{\tau}_1\cdot\boldsymbol{\tau}_2)\nonumber\\
&+&C_5(\boldsymbol{\sigma}_1\cdot\boldsymbol{q})(\boldsymbol{\sigma}_2\cdot\boldsymbol{q})+C_6(\boldsymbol{\sigma}_1\cdot\boldsymbol{q})(\boldsymbol{\sigma}_2\cdot\boldsymbol{q})(\boldsymbol{\tau}_1\cdot\boldsymbol{\tau}_2)\nonumber\\
&+& \frac{i}{2}C_7(\boldsymbol{q}\times\boldsymbol{k})\cdot(\boldsymbol{\sigma}_1+\boldsymbol{\sigma}_2),
\label{eq:2Ncontact}
\end{eqnarray}
where $B_i$ and $C_i$ are  LECs fitted to the Nijmegen neutron-proton ($n$-$p$) scattering phase shifts~\cite{Stoks:1993}. 
All contact terms are regularized using a nonlocal Gaussian regulator in  momentum space, following Ref.~\cite{Lu:2023b}. 
Specifically, for the matrix element $ V(\boldsymbol{p}_1,\boldsymbol{p}_1', \boldsymbol{p}_2,\boldsymbol{p}_2')$, where $\boldsymbol{p}_i$ and $\boldsymbol{p}'_i$ are the  incoming and outgoing momenta 
of the $i$th nucleon, we apply a non-local momentum regulator
\begin{equation} 
f_{2\mathrm{N}}=\prod_{i=1}^2 \exp\left(-\frac{p_i^6+p_i'^6}{2\Lambda^6}\right).
\label{eq:f2N}
\end{equation}
In the center-of-mass frame, Eq.~(\ref{eq:f2N}) reduces to the nonlocal regulator commonly used in Ref. ~\cite{Entem:2003}.

The isospin-breaking term, $V_{\mathrm{IB}}$, first appears at NLO. Following the convention of Ref.~\cite{PhysRevLett.104.142501}, we separate it into three components: the long-range Coulomb term,
\begin{equation}  
\label{eq:Vcou}
V_{\mathrm{IB}}^{\mathrm{Cou}} = \frac{\alpha_{\mathrm{EM}}}{\mathrm{max}(|\boldsymbol{r}_1-\boldsymbol{r}_2|,0.5)}\left(\frac{1+\tau_{1,z}}{2}\right)\left(\frac{1+\tau_{2,z}}{2}\right),
\end{equation}
the short-range proton-proton ($p$-$p$) contact term,
\begin{equation}  
\label{eq:Vpp}
V_{pp}=C_{pp}\delta(\boldsymbol{r}_1-\boldsymbol{r}_2)\left(\frac{1+\tau_{1,z}}{2}\right)\left(\frac{1+\tau_{2,z}}{2}\right),
\label{eq:pp}
\end{equation}
and the short-range neutron-neutron ($n$-$n$) contact term,
\begin{equation}
\label{eq:Vnn}
V_{nn}=C_{nn}\delta(\boldsymbol{r}_1-\boldsymbol{r}_2)\left(\frac{1-\tau_{1,z}}{2}\right)\left(\frac{1-\tau_{2,z}}{2}\right).
\label{eq:nn}
\end{equation}
These isospin-breaking terms are essential for reproducing the $^3$H-$^3$He mass splitting~\cite{Epelbaum:2010}. 
The 2N LECs $C_{pp}$ and $C_{nn}$ in Eqs.~(\ref{eq:pp}) and (\ref{eq:nn}) are fitted to 
experimental $p$-$p$ phase shifts, as well as $n$-$n$ scattering lengths and effective ranges~\cite{Trotter:2006,Miller:1990}. Details on determining the LECs are provided in Appendix~\ref{app:a1}.

Finally, for the three-nucleon (3N) potential $V_{3\mathrm{N}}^{Q^3}$, which first appears at N$^2$LO,  we include the OPE term
\begin{eqnarray}
V_{3\mathrm{N}}^{\mathrm{OPE}}&=&-\frac{g_A}{8f^2_\pi} \frac{c_D}{f_\pi^2\Lambda_\chi}\sum_{C(i,j,k)}\frac{(\boldsymbol{\sigma}_i\cdot\boldsymbol{q}_i)(\boldsymbol{\sigma}_j\cdot\boldsymbol{q}_i)}{\boldsymbol{q}_i^2+M_\pi^2}(\boldsymbol{\tau}_i\cdot\boldsymbol{\tau}_j),\nonumber\\
&&
\end{eqnarray}
and the 3N contact term
\begin{equation}   \label{eq:V3NCT}V_{3\mathrm{N}}^{\mathrm{CT}}=\frac{c_E}{2f_\pi^4\Lambda_\chi} \sum_{C(i,j,k)}\boldsymbol{\tau}_i\cdot\boldsymbol{\tau}_j.
\end{equation}
Here, $\Lambda_\chi=700$ MeV is the chiral symmetry breaking scale. The 3N regulator
\begin{equation}
  f_{3\mathrm{N}}=\prod_{i=1}^3 \exp\left(-\frac{p_i^6+p_i'^6}{2\Lambda^6}\right)  
\end{equation}
is employed to regularize $V_{3\mathrm{N}}^{\mathrm{CT}}$.  We have also tested the 3N TPE potential but found its contribution negligible, 
so we neglect it in this work. The LECs $c_D$ and $c_E$ are determined by fitting to the experimental $^3$H binding energy and  the RGTME of $^3$H $\beta$-decay process simultaneously. Details are given in  Sec.~\ref{sec:Res}.

\subsection{Nonperturbative interaction}
\label{subsec:LO interaction}

The full Hamiltonian  in Eq.~(\ref{eq:fullH}) includes complex spin-isospin-dependent terms, which pose significant challenges for lattice calculations. One major issue is that these terms can induce severe sign problems in fermionic systems, such as nucleons, in Monte Carlo simulations~\cite{Lahde:2019}. To mitigate this, one can construct a simplified Hamiltonian $H_0$ which captures the essential physics but with much milder sign problems, and solve the full Hamiltonian $H$  perturbatively~\cite{Lu:2021}. 
Details of the perturbative method will be provided in Sec.~\ref{subsec:Monte Carlo}.
In this work, we consider the following nonperturbative Hamiltonian:
\begin{eqnarray}
    H_{0}& = &K + \frac{C_{2}}{2!}\sum_{\boldsymbol{n}}:\tilde{\rho}^2(\boldsymbol{n}): + \frac{C_3}{3!}\sum_{\boldsymbol{n}}:\tilde{\rho}^3(\boldsymbol{n}):\nonumber\\
    &+&\frac{C_I}{2!}\sum_{I,\boldsymbol{n}}:\tilde{\rho}^2_I(\boldsymbol{n}):+V_{\mathrm{OPE}}^{\Lambda_\pi'}.
    \label{eq:LOH}
\end{eqnarray}
Here, :: represents normal ordering and $\tilde{\rho}$ is the SU(4)-symmetric density operator smeared both locally and nonlocally~\cite{Elhatisari:2017}:
\begin{eqnarray}
   \tilde{\rho}(\boldsymbol{n}) &=& \sum_{s,i}\tilde{a}^\dagger_{s,i}(\boldsymbol{n})\tilde{a}_{s,i}(\boldsymbol{n})\nonumber\\
   &+&s_{\mathrm{L}}\sum_{|\boldsymbol{n}-\boldsymbol{n}'|=1}\sum_{s,i}\tilde{a}^\dagger_{s,i}(\boldsymbol{n}')\tilde{a}_{s,i}(\boldsymbol{n}'),
\end{eqnarray}
where $s$ and $i$ are the spin and isospin indexes, respectively. The parameter $s_{\mathrm{L}}$ controls the range of
the local part of the interaction. The smeared creation operator is defined as
\begin{equation} \tilde{a}^\dagger_{s,i}(\boldsymbol{n}) = a^\dagger_{s,i}(\boldsymbol{n})+s_{\mathrm{NL}}\sum_{|\boldsymbol{n}-\boldsymbol{n}'|=1}\tilde{a}^\dagger_{s,i}(\boldsymbol{n}'),
\end{equation}
with $s_{\mathrm{NL}}$ controlling the range of the nonlocal part of the interaction.
The definition of the isospin-dependent density operator $\tilde{\rho}_I$ is similar to $\tilde{\rho}$:
\begin{eqnarray}
   \tilde{\rho}_I(\boldsymbol{n}) &= &\sum_s\sum_{i,i'}\tilde{a}^\dagger_{s,i}(\boldsymbol{n})[\tau_I]_{i,i'}\tilde{a}_{s,i'}(\boldsymbol{n})\nonumber\\
&+&s_{\mathrm{L}}\sum_{|\boldsymbol{n}-\boldsymbol{n}'|=1}\sum_s\sum_{i,i'}\tilde{a}^\dagger_{s',i}(\boldsymbol{n}')[\tau_I]_{i,i'}\tilde{a}_{s',i'}(\boldsymbol{n}'),\nonumber\\  
\end{eqnarray}
with $\tau_I$ the $I$th Pauli matrix .

The Hamiltonian in Eq.~(\ref{eq:LOH}) consists of two parts, written separately for clarity. The first line contains terms that respect Wigner's SU(4) symmetry~\cite{Wigner:1937}, an approximate symmetry of low-energy nuclear interactions. The inclusion of this symmetry helps suppress sign problems in lattice Monte Carlo calculations~\cite{Chen:2004}.
%and enables a practical extraction of $\beta$-decay matrix elements. 
Recently, the SU(4)-symmetric Hamiltonian has been applied to the study of the properties of light and medium-mass nuclei as well as the equation of state of symmetric nuclear matter and neutron matter~\cite{Lu:2019}, the intrinsic clustering geometry of $^{12}$C's low-lying states~\cite{Shen:2023}, the thermodynamics of nuclear systems~\cite{Lu:2020}, the clustering effects in hot, dilute nuclear matter~\cite{Ren:2023}, and the $\alpha$-particle monopole transition form factor~\cite{Ulf:2024}.
The second line of Eq.~(\ref{eq:LOH}) introduces an isospin-dependent contact term and a OPE term, both of which explicitly break the SU(4) symmetry. 
These additional terms are introduced to improve the overall description of $p$-shell nuclei~\cite{Wiringa:2002} , such as $^6$He and $^6$Li studied in this work. To suppress the sign problem caused by the tensor force,  we adopt a lower cutoff of $\Lambda_\pi' = 180$ MeV for the OPE term in Eq.~(\ref{eq:LOH}), compared to the higher cutoff of $\Lambda_\pi = 300$ MeV used in the full Hamiltonian in Eq.~(\ref{eq:fullH}).

To determine the unknown parameters in Eq.~(\ref{eq:LOH}), we perform a combined fit to $n$-$p$ scattering phase shifts in two $S$-wave channels, 
along with the ground-state binding energies of $^4$He and $^6$He. The inclusion of the latter two observables is motivated by the fact that $^6$He 
is a weakly bound halo nucleus~\cite{Tanihata:1995,Jensen:2004,Zhukov:1993}, with its binding energy very close to that of $^4$He:
\begin{equation}   
E(^4\mathrm{He}) = -28.30\mathrm{MeV},\ E(^6\mathrm{He}) = -29.27\mathrm{MeV}.
\end{equation}
Therefore, we carefully tune $H_0$ to ensure that $^6$He is bound in our calculation. The fitted parameter values are presented in Table~\ref{tab:table0}.
\begin{table}[H]
\centering
\renewcommand{\arraystretch}{1.2}
\resizebox{230pt}{!}
		{\begin{tabular}{c c c c c}
			\hline\hline
			     $s_{\mathrm{L}}$& $s_{\mathrm{NL}}$& $C_2(\mathrm{MeV}^{-2})$& $C_I(\mathrm{MeV}^{-2})$& $C_3(\mathrm{MeV}^{-5})$\\
			\hline
			0.109&   0.122 &-6.41$\times$10$^{-6}$ &9.20$\times$10$^{-7}$&-9.65$\times$10$^{-13}$\\
				\hline\hline
		\end{tabular}}
		\caption{Values for all the parameters in Eq.~(\ref{eq:LOH}).}
		\label{tab:table0}
\end{table}

\section{Axial Current}
\label{sec:current}

The transition operator for the RGTME is related to the nuclear axial current. We employ one- and two-body $\Delta$-less chiral EFT axial currents derived from the unitary transformation method of the Bonn group~\cite{Krebs:2017}  and the time-ordered perturbation method of the JLab-Pisa group~\cite{Girlanda:2016} . 
These two approaches yield consistent tree-level results up to N$^3$LO ($Q^0$) but exhibit some differences at N$^4$LO ($Q$) due to loop contributions~\cite{Baroni:2017,Krebs:2017,Krebs:2020}.
Moreover, Ref.~\cite{Krebs2:2020} recently highlighted that the absence of consistent regularization for axial-vector current operators leads to a violation of chiral symmetry in the chiral limit at N$^4$LO. For these reasons, we restrict our analysis to tree-level axial currents up to N$^3$LO and neglect higher-order contributions.

\begin{figure}[htbh]
\includegraphics[height=1.0in]{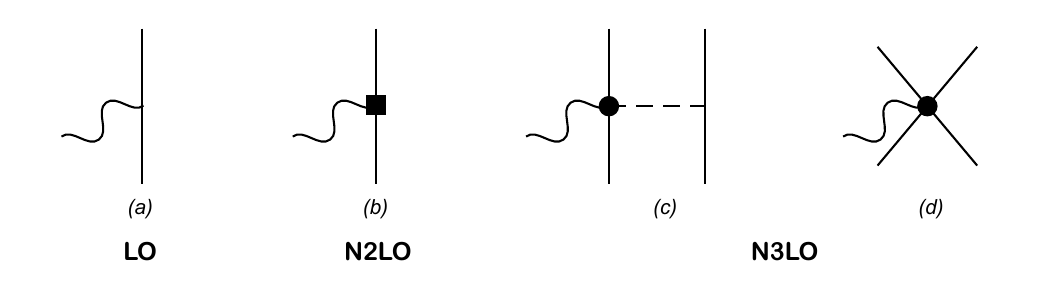}
\caption{Diagrams illustrating the contributions to the axial
current up to N$^3$LO used in this work. Nucleons, 
pions and the external field are denoted by the solid ,
dashed, and wavy lines respectively. Panel (a) denotes the LO current while panel (b) is the relativistic correction at N$^2$LO. Panels (c) and (d) are the OPE current and the contact current at N$^3$LO, respectively. }
\label{fig:current}
\end{figure}

Since the momentum $\boldsymbol{k}$ carried by the current is very small in $\beta$-decay processes, we work in the  $\boldsymbol{k}=0$ limit  to simplify calculations. Under this limit, there are in total four  terms that contribute to the axial current up to N$^3$LO: the one-body current $\boldsymbol{A}_{\mathrm{LO}}$ at LO , the relativistic correction term $\boldsymbol{A}_{\mathrm{N}^2\mathrm{LO}}$ at N$^2$LO,  the OPE term $\boldsymbol{A}_{\mathrm{N}^3\mathrm{LO}}(\mathrm{OPE})$, and the contact term $\boldsymbol{A}_{\mathrm{N}^3\mathrm{LO}}(\mathrm{CT})$  at N$^3$LO.  Figure~\ref{fig:current} provides a diagrammatic illustration of them, and their expressions in the momentum space can be found in Refs.~\cite{Baroni:2017, Park:2003}.   
 The OPE current involves two nucleon-pion coupling constants, $c_3$ and $c_4$, for which we take the values $c_3 = -5.61$ GeV$^{-1}$ and $c_4 = 4.26$ GeV$^{-1}$~\cite{Hoferichter:2015}. The contact current is related to the three-body force coefficient $c_D$ as follows~\cite{Park:2003, Baroni:2018}:
 \begin{equation}
\boldsymbol{A}_{\mathrm{N}^3\mathrm{LO}}(\mathrm{CT})=-\frac{c_D}{8f_\pi^2\Lambda_\chi}  \sum_{i<j}^A (\boldsymbol{\tau}_i\times\boldsymbol{\tau}_j)^+(\boldsymbol{\sigma}_i\times\boldsymbol{\sigma}_j) ,
 \end{equation}
 where we used the shorthand notation $(\boldsymbol{\tau}_i\times\boldsymbol{\tau}_j)^+=(\boldsymbol{\tau}_i\times\boldsymbol{\tau}_j)^x +i(\boldsymbol{\tau}_i\times\boldsymbol{\tau}_j)^y$ for the two-body isospin-raising operator.

 The realization of the axial current on the lattice is straightforward. Given the momentum-space expression of the current, the finite lattice spacing and box size provides natural ultraviolet and infrared cutoffs, respectively, which are used to discretize the momenta and constrain them in the first Brillouin zone. The corresponding operator in the configuration space is then constructed through fast Fourier transformation. 

As a comparison, in the following  we briefly discuss the treatment of two-body currents for calculations in the continuum~\cite{Park:2003,Gazit:2009, Baroni:2018, King:2020, Gnech:2021}. These calculations rely on analytical  configuration-space expressions of the OPE and  contact currents,  which look like
 \begin{eqnarray}    \label{eq:xxx}  \boldsymbol{A}_{\mathrm{N}^3\mathrm{LO}}(\mathrm{OPE})&= & \frac{g_A}{2f_\pi^2}\left(\frac{1}{3}c_3+\frac{2}{3}c_4+\frac{1}{6M_N}\right)\sum_{i<j}^A\delta^{(3)}(\boldsymbol{r}_{ij})\nonumber\\ 
     &&\times(\boldsymbol{\tau}_i\times\boldsymbol{\tau}_j)^+(\boldsymbol{\sigma}_i\times \boldsymbol{\sigma}_j)+\tilde{\boldsymbol{A}}_{\mathrm{N}^3\mathrm{LO}}(\mathrm{OPE})\nonumber\\   \boldsymbol{A}_{\mathrm{N}^3\mathrm{LO}}(\mathrm{CT})&= &-\frac{c_D}{8f_\pi^2\Lambda_\chi}  \sum_{i<j}^A\delta^{(3)}(\boldsymbol{r}_{ij}) (\boldsymbol{\tau}_i\times\boldsymbol{\tau}_j)^+(\boldsymbol{\sigma}_i\times\boldsymbol{\sigma}_j).\nonumber\\
     &&
    \end{eqnarray}
    The above equations can be easily derived by performing Fourier transformation on momentum-space expressions of the current. Note that $\boldsymbol{A}_{\mathrm{N}^3\mathrm{LO}}(\mathrm{OPE})$ consists of two parts: the first part proportional to $\delta^{(3)}(\boldsymbol{r}_{ij})= \delta^{(3)}(\boldsymbol{r}_{i}-\boldsymbol{r}_j)$ absorbs the short-range singularity,  while the second part with a tilde $\tilde{\boldsymbol{A}}_{\mathrm{N}^3\mathrm{LO}}$(OPE) collects remaining long-range contributions, whose explicit form can be found in Refs.~\cite{Pastore:2018, King:2020, Gnech:2021}. Because of the Fermi-Dirac statistics~\cite{Park:2003}, the first part has the same spin-isopsin structure as the contact current $\boldsymbol{A}_{\mathrm{N}^3\mathrm{LO}}(\mathrm{CT})$. Therefore, some works absorb the former into the latter and call the remaining term $\tilde{\boldsymbol{A}}_{\mathrm{N}^3\mathrm{LO}}$(OPE) as the OPE current~\cite{Park:2003,Gazit:2009, Baroni:2018, King:2020, Gnech:2021}. There is also literature which defines the full operator $\boldsymbol{A}_{\mathrm{N}^3\mathrm{LO}}$(OPE) as the OPE current~\cite{Baroni:2017}. We mention that these two definitions are equivalent to each other, and would lead to consistent predictions on physical observables such as the half-life. For our lattice calculation, we find it more  convenient to follow the second prescription, which makes it easier to transform the current between the momentum representation and configuration representation, numerically.

\section {Computational Methods}
\label{sec:cal method}

In this section, we introduce two  methods used in this work, i.e., the Monte Carlo method and the sparse-matrix diagonalization method, to calculate the RGTME following the convention of Edmonds~\cite{Edmonds:1955},
\begin{equation}
    \mathrm{GT}= \frac{\sqrt{2J_f+1}}{g_A}\frac{\langle J_fM_f|\mathcal{J}|J_iM_i\rangle}{\langle J_iM_i, 11|J_fM_f\rangle}.
    \label{eq:GTdef}
\end{equation}
Here, $|J_{i}M_{i}\rangle$ and  $|J_{f}M_{f}\rangle$ represent the eigenstates of the full Hamiltonian $H$ for the initial and final nucleus, respectively, with $J$ the nuclear spin and $M$ the magnetic quantum number.  
The weak transition operator $\mathcal{J}$ is defined as 
\begin{equation}   \mathcal{J}=-\frac{1}{\sqrt{2}}(A_{x}+iA_{y}),
\end{equation}
with $A_{x,y}$ the sum of axial currents up to N$^3$LO. Since $\mathcal{J}$ is a rank-1 spherical tensor carrying the magnetic quantum number $m_{\mathcal{J}}=1$, the Clebsch-Gordan coefficient $\langle J_iM_i, 11|J_fM_f\rangle$ is required to appear in the denominator to extract the reduced matrix element.  For the processes studied in this paper, we have $J_i=J_f=\frac{1}{2}$ for $^3$H$\rightarrow^3$He $\beta$ decay and $J_i=0, J_f = 1$ for $^6$He$\rightarrow^6$Li $\beta$ decay, and we set $M_i = -1/2$, $M_f= 1/ 2$ for the former transition and $M_i=0$, $M_f = 1$ for the latter. 

We comment that the definition of the RGTME, Eq.~(\ref{eq:GTdef}),
is followed throughout the work. The same convention has also been  adopted to study $^{6}$He $\beta$ decay in Ref.~\cite{Gnech:2021,King:2020}, so our result of $^{6}$He's RGTME can be directly compared to theirs. For $^3$H $\beta$ decay, however, different conventions have been adopted by different works. For example, our definition of the RGTME is larger than the Gamow-Teller matrix element defined in Ref.~\cite{Baroni:2017, Baroni:2018,Gnech:2021,Elhatisari:2024gg} by a factor of $\sqrt{2}$, but the same as the one defined in Ref.~\cite{SU4_and_beta_decay}. One should be aware of this difference when making a comparison.

For latter convenience, we decompose $\mathcal{J}$ into different orders according to Fig.~\ref{fig:current},
\begin{equation}
    \mathcal{J} =\mathcal{J}_{\mathrm{LO}}+\mathcal{J}_{\mathrm{N}^2\mathrm{LO}}+\mathcal{J}_{\mathrm{N}^3\mathrm{LO}}.
\end{equation}
Accordingly, the RGTME can be expanded order by order as
\begin{equation}    \mathrm{GT}=\mathrm{GT}_{\mathrm{LO}}+\mathrm{GT}_{\mathrm{N}^2\mathrm{LO}}+\mathrm{GT}_{\mathrm{N}^3\mathrm{LO}}.
\end{equation}

\subsection{Monte Carlo method}
\label{subsec:Monte Carlo}

The Monte Carlo method is a crucial tool for simulating nuclear systems in NLEFT. In this section, we discuss its application to our work.

We begin by introducing the full transfer matrix
\begin{equation}
    M = :e^{-a_t H}:
\label{eq:transfermatrix}
\end{equation}
for the full Hamiltonian $H$ defined in Eq.~(\ref{eq:fullH}), where $a_t$ is the temporal lattice spacing. 
To facilitate perturbative calculations that will be introduced below, we also define a nonperturbative transfer matrix
\begin{equation}
    M_0 = :e^{-a_t H_{0}}:,
\end{equation}
corresponding to the nonperturbative Hamiltonian $H_{0}$ defined in Eq.~(\ref{eq:LOH}).

Next, we prepare the trial states $|\Phi_{i/f}\rangle$, which share the same quantum numbers as $|J_{i/f}M_{i/f}\rangle$. These states are evolved along Euclidean time using the transfer matrix $M$, 
\begin{equation}   
|\Phi^{t/2}_{i/f}\rangle\equiv M^{N_t/2}|\Phi_{i/f}\rangle,
\end{equation}
where $N_t$ is the number of evolution steps and $t=N_ta_t$ represents the evolution time. Since both processes studied here, i.e., $^3\mathrm{H}\rightarrow ^3\mathrm{He}$ and $^6\mathrm{He}\rightarrow $\ $^6\mathrm{Li}$, only involve ground-state nuclei, $t$ is taken as large as possible to suppress excited-state contamination. To accelerate the convergence, the trial states $|\Psi_{i/f}\rangle$ are constructed based on the the nuclear shell model~\cite{Shellmodel_Mayer,Shellmodel_Heyde}.

 The RGTME, given in Eq.~(\ref{eq:GTdef}), can be extracted from the ratio
\begin{equation}
    \mathrm{GT}(t) =\frac{\sqrt{2J_f+1}}{g_A\langle J_iM_i, 11|J_fM_f\rangle}\frac{\langle \Phi^{t/2}_f|\mathcal{J} |\Phi^{t/2}_i\rangle }{\sqrt{\langle \Phi^{t/2}_i|\Phi^{t/2}_i\rangle \langle \Phi^{t/2}_f|\Phi^{t/2}_f\rangle}}
\label{eq:GTlatdef}
    \end{equation}
at large $t$. 
To evaluate it, auxiliary field transformations are applied to both the numerator and denominator, reducing the problem to a multidimensional integral that can be handled using Monte Carlo methods. A detailed discussion of this process is provided in Appendix~\ref{app:a4}. However, as discussed in Sec.~\ref{subsec:LO interaction}, a direct auxiliary field transformation on the full transfer matrix $M$ leads to severe sign problems, making the results highly noisy.

To address this issue, we employ perturbation theory~\cite{Lu:2021}. Specifically, we decompose the transfer matrix into
\begin{equation}
    M=M_{0}+\Delta M,
    \label{eq: split}
\end{equation}
treating $\Delta M$ as a perturbation. This allows us to expand Eq.~(\ref{eq:GTlatdef}) as a perturbative series:
\begin{eqnarray}
    \mathrm{GT}=\sum_{n=0}^{\infty} \mathrm{GT}^{(n)}_{\mathrm{LO}} + \sum_{n=0}^{\infty} \mathrm{GT}^{(n)}_{\mathrm{N}^2\mathrm{LO}}+\sum_{n=0}^{\infty}\mathrm{GT}^{(n)}_{\mathrm{N}^3\mathrm{LO}},
    \label{eq:PTseries}
\end{eqnarray}
with the time variable $t$ suppressed for brevity. Here, $n$ denotes the order of expansion in $\Delta M$. The contributions from different orders of the current are written as LO, N$^2$LO, and N$^3$LO, respectively. For the LO current, we retain the first two terms in the perturbative series and neglect higher-order corrections, namely,
\begin{equation}
\mathrm{GT}_{\mathrm{LO}}\approx \mathrm{GT}^{(0)}_{\mathrm{LO}}+\mathrm{GT}^{(1)}_{\mathrm{LO}}.
\label{eq:PTLO}
\end{equation}
For currents beyond LO, we keep only the first term,
\begin{equation}   \mathrm{GT}_{\mathrm{N}^{2}\mathrm{LO}/\mathrm{N}^{3}\mathrm{LO}}\approx\mathrm{GT}^{(0)}_{\mathrm{N}^{2}\mathrm{LO}/\mathrm{N}^{3}\mathrm{LO}}.
\label{eq:PTN2LO}
\end{equation}
Thus, Eq.~(\ref{eq:PTseries}) is approximated as 
\begin{eqnarray} \mathrm{GT}&\approx&\sum_{n=0}^1\mathrm{GT}^{(n)}_{\mathrm{LO}}+\mathrm{GT}^{(0)}_{\mathrm{N}^{2}\mathrm{LO}}+\mathrm{GT}^{(0)}_{\mathrm{N}^{3}\mathrm{LO}}.
\label{eq:GTfinal}
\end{eqnarray}
The explicit expressions for the terms on the right-hand side of Eq.~(\ref{eq:GTfinal}) are provided in Appendix~\ref{app:a3}. 
They are computed using Monte Carlo methods. Since the system now evolves under $H_0$ that has an approximate Wigner-SU(4) symmetry 
instead of the full Hamiltonian $H$, the sign problem is significantly alleviated.

When designing the algorithm, we note that any $\beta$-decay process involves two different nuclei, making this essentially a two-channel problem. 
To control statistical errors, an appropriate probability distribution function is required to update the auxiliary fields during the Markov process. We leave relevant details to  Appendix~\ref{app:a4}.

\subsection{Lanczos eigenvector method}
For the $^3$H$\rightarrow{^3}$He process, the dimension of the Hilbert space is not too large and a rigorous calculation is possible. Therefore, in addition to the Monte Carlo approach, we employ the Lanczos eigenvector technique as an alternative method~\cite{Epelbaum:2010}.
Specifically, we directly diagonalize the full Hamiltonian $H$ to obtain the ground-state wave functions of $^3$H and $^3$He. These wave functions are then used as input to compute the RGTME following Eq.~(\ref{eq:GTdef}). We make the calculation on a box with its length $L\approx 20$ fm, which is sufficiently large to neglect finite-volume effects~\cite{Elhatisari:2024gg}.

The main advantage of the Lanczos method is its accuracy. Unlike Monte Carlo simulations, it is free from statistical errors and systematic errors from the perturbative method. Additionally, due to its precision, this method enables a precise determination of the three-body LECs $c_D$ and $c_E$ from $^3$H $\beta$ decay.
Specifically, we fit them to the experimental binding energy and half-life of $^3$H, closely following the procedure in  Refs.~\cite{Gazit:2009, Baroni:2017, Baroni:2018,Elhatisari:2024gg}. The details are presented in the next section.

\section{Results}
\label{sec:Res}

\subsection{$^3$H $\beta$ decay}
The details of $^3$H $\beta$ decay are discussed in this subsection.  An essential step here is to convert the half-life, $t_{1/2}$, into the RGTME. We make use of the formula~\cite{Simpson:1987,Hardy:1990,Schiavilla:1998,Chou:1993}
\begin{equation}
  (1+\delta_R)t_{1/2}f_V = \frac{2\pi^3\mathrm{ln}2/[m_e^5 V_{ud}^2 G_F^2 (1+\Delta_R^V)]}{ B(\mathrm{F})+ B(\mathrm{GT})f_A/f_V},
    \label{eq:half_life}
\end{equation} 
which expresses $t_{1/2}$ in terms of the Fermi transition strength
\begin{equation}
B(\mathrm{F})=\left(\sum_{n=1}^3\langle   ^3\mathrm{He}|\tau_{n,+}|^3\mathrm{H}\rangle\right)^2,
\end{equation}
and the Gamow-Teller transition strength $B(\mathrm{GT})$, quadratic in the RGTME~\cite{Chou:1993},
\begin{equation}
    B(\mathrm{GT}) = \frac{1}{2}g_A^2 \mathrm{GT}^2.
\end{equation}
 In Eq.~(\ref{eq:half_life}), $m_e=0.511$ MeV is the electron mass, $V_{ud}=0.9737(3)$ is the up-down quark-mixing element of the Cabibbo-Kobayashi-Maskawa matrix~\cite{Hardy:2020}, $G_F=1.1664\times10^{-5}$ GeV$^{-1}$ is the Fermi constant~\cite{MuLan:2012}, $f_V=2.8355\times10^{-6}$ and $f_A=2.8505\times10^{-6}$ are Fermi functions~\cite{Simpson:1987}, $\delta_R=1.9\%$ is the outer radiative correction of $^3$H $\beta$ decay \cite{Raman:1978}, and $\Delta_R^V=0.02467(22)$ is the inner radiative correction~\cite{Seng:2018}.  Our lattice result of the Fermi transition strength is
\begin{equation}
\label{eq:Fermi Strength}
   B(\mathrm{F}) = 0.9996^2.
\end{equation}
Note that because of vector current conservation, $B(\mathrm{F})$ is strictly equal to 1 in the isospin-symmetric limit. In our calculation, the isospin-breaking potential $V_{\mathrm{IB}}$ introduced in Eqs.~(\ref{eq:Vcou}),   
 (\ref{eq:Vpp}) and (\ref{eq:Vnn}) generates a small shift, at the level of $10^{-4}$. Combining the Fermi transition strength Eq.~(\ref{eq:Fermi Strength}) with the measured comparative half-life, $(1+\delta_R)t_{1/2}f_V=1134.6(31)$  s~\cite{Simpson:1987},  the value of the RGTME is determined as
\begin{equation} \mathrm{GT}_{\mathrm{exp}} = 2.3297(58).
\end{equation}
 The subscript “$\mathrm{exp}$” means that the result is derived from experimental data, although the lattice prediction of the Fermi transition strength is used as an input. Note that our definition of $^3$H's RGTME is different from Ref.~\cite{Gazit:2009, Baroni:2017, Baroni:2018,Elhatisari:2024gg} by an overall constant $\sqrt{2}$, as has been emphasized in the last section.

\begin{figure}
\includegraphics[height=2.4in]{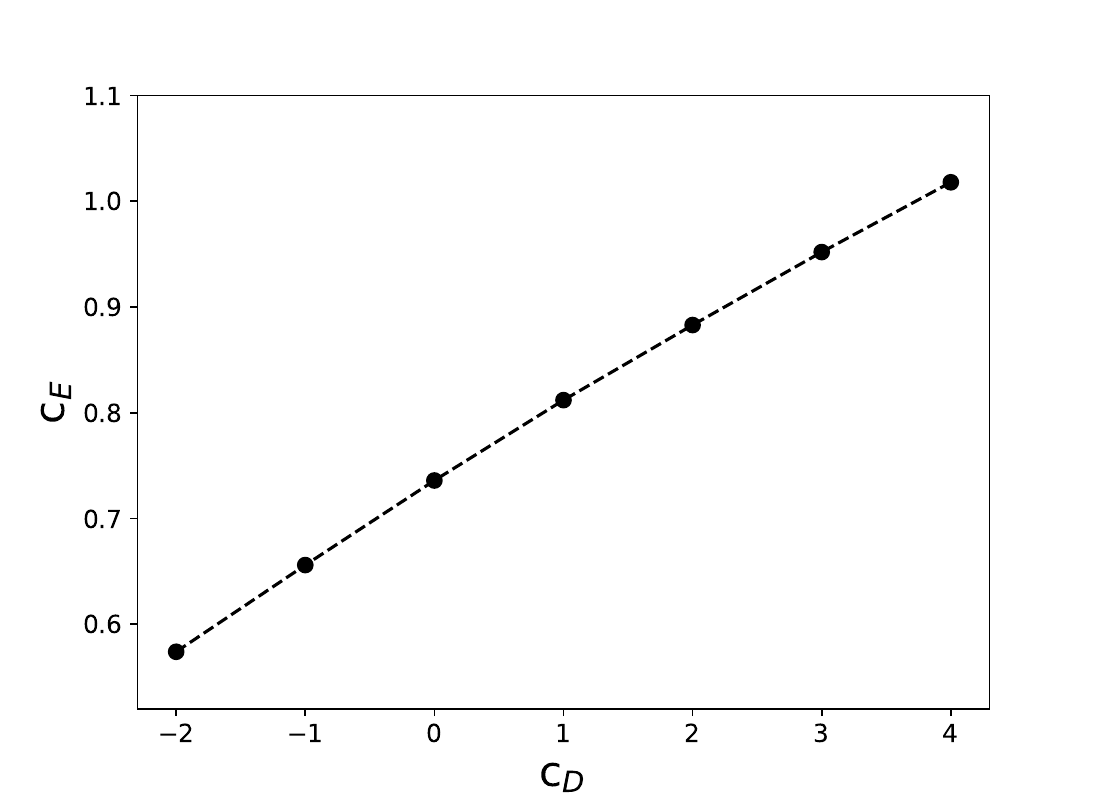}
\caption{The $c_E$-$c_D$ trajectories obtained by
fitting the experimental $^3$H binding energy 8.482 MeV.}
\label{fig:cDandcE}
\end{figure}

\begin{figure}
\includegraphics[height=2.4in]{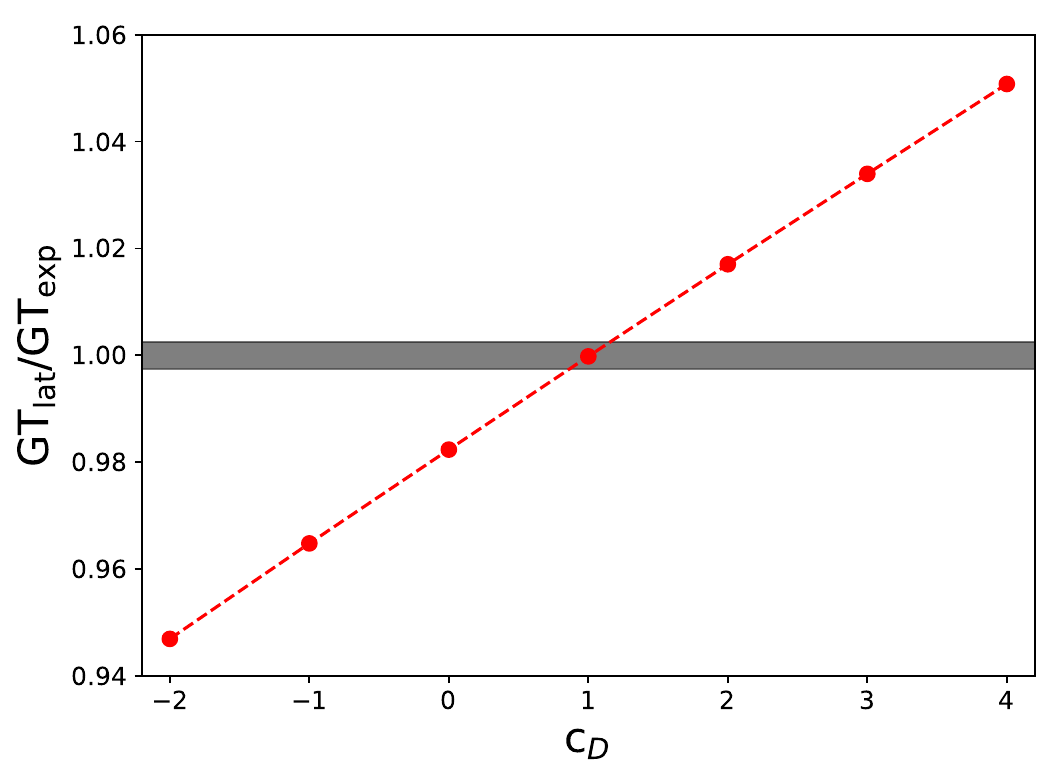}
\caption{The  ratio GT$_{\mathrm{lat}}$/GT$_{\mathrm{exp}}$ as a function
of $c_D$. The gray band shows the experimental uncertainty of GT$_{\mathrm{exp}}$.}
\label{fig:GTandcD}
\end{figure}

Now we show the determination of $c_D$ and $c_E$ using the Lanczos eigenvector method. In Fig.~\ref{fig:cDandcE}, $c_E$ is plotted against $c_D$ by fitting to the experimental binding energy $E(^3$H$)=-8.482$ MeV. Similar to previous studies, 
we observe a nearly linear relation between $c_D$ and $c_E$~\cite{Gazit:2009, Baroni:2017, Baroni:2018}. Then we calculate the lattice RGTME, $\mathrm{GT}_{\mathrm{lat}}$,  and plot its ratio with respect to $\mathrm{GT}_{\mathrm{exp}}$ in Fig.~\ref{fig:GTandcD}, as a function of $c_D$. All the points in Fig.~\ref{fig:GTandcD} share the same values of $(c_D, c_E)$ with the ones in Fig.~\ref{fig:cDandcE}. Again, GT$_{\mathrm{th}}$/GT$_{\mathrm{exp}}$ is a nearly linear function of $c_D$, and the values of $c_D$ and $c_E$ are determined as
\begin{equation}
    c_D = 1.01(16), \quad c_E = 0.81(1).
    \label{eq:3Nvalue}
\end{equation}
Note that low-energy nucleon-deuteron ($n$-$d$) scattering provides an alternative process to help pin down $c_D$ and $c_E$ and to test the result above  from $^3$H $\beta$ decay. The strategy has been employed by several works in the continuum ~\cite{nd_scattering:2002, nd_scattering:2018, nd_scattering:2019}, 
but it should be understood
 that a direct comparison of $c_D$ and $c_E$ cannot be performed due to different regularization schemes and regulators 
 used in our work and in ~\cite{nd_scattering:2002, nd_scattering:2018, nd_scattering:2019}. In NLEFT, the adiabatic projection method ~\cite{adiabatic_projection1, adiabatic_projection2} was applied to investigating $n$-$d$ scattering~\cite{nd_scattering_LEFT1, nd_scattering_LEFT2} using LO pionless EFT, with no discussion on the determination of three-body forces. Thus, a more systematic study of  $n$-$d$ scattering on the lattice is required.
 
\begin{center}
	\begin{table}[H]
    \renewcommand\arraystretch{1.2}
	\resizebox{250pt}{!}{
		\begin{tabular}{l c c c c}
			\hline\hline
			       & LO         &        N$^2$LO   &      N$^3$LO(OPE) & N$^3$LO(CT)\\
			\hline
			LE &   2.3684   & -0.0117&   -0.0679   & 0.0409 \\
                MC &  2.355(17) &
                -0.010(2) & -0.070(1) & 0.039(3)\\
		\hline\hline
		\end{tabular}
		}
		\caption{Contributions to RGTME from currents at different orders for $c_D=$1.01, $c_E$=0.81. The upper and lower lines  are the results of the Lanczos eigenvector method (LE) and the Monte Carlo (MC) method, respectively. The value in parentheses is the statistical error.}
		\label{tab:table1}
	\end{table}
\end{center}
 
                                \begin{center}
\begin{table}[H]
\renewcommand{\arraystretch}{1.2}
   \resizebox{250pt}{!}{ \begin{tabular}{c| c |c| c| c}
    \hline\hline
         Nuclei($J^P$)& \multicolumn{2}{c|}{$^6$He(0$^+$)}&\multicolumn{2}{c}{$^6$Li(1$^+$)}\\
          \hline
         & $E$/MeV&$R_{\mathrm{ch}}$/fm& $E$/MeV&$R_{\mathrm{ch}}$/fm \\
         \hline
         Lattice&-28.80(1.21)&1.988(32)&-31.88(93)&2.534(18)\\
         \hline      Expt.&-29.27&2.068(11)&-31.99&2.589(39)\\
    \hline\hline
    \end{tabular}}
    \caption{ Ground state energies and the charge radii of $^6$He and $^6$Li calculated on the lattice compared to experiment. The third line is the lattice prediction and the last line is the experimental result.}
    \label{tab:He6Li6observable}
\end{table}
\end{center}

\begin{figure*}
\centering
\includegraphics[height=3.2in]{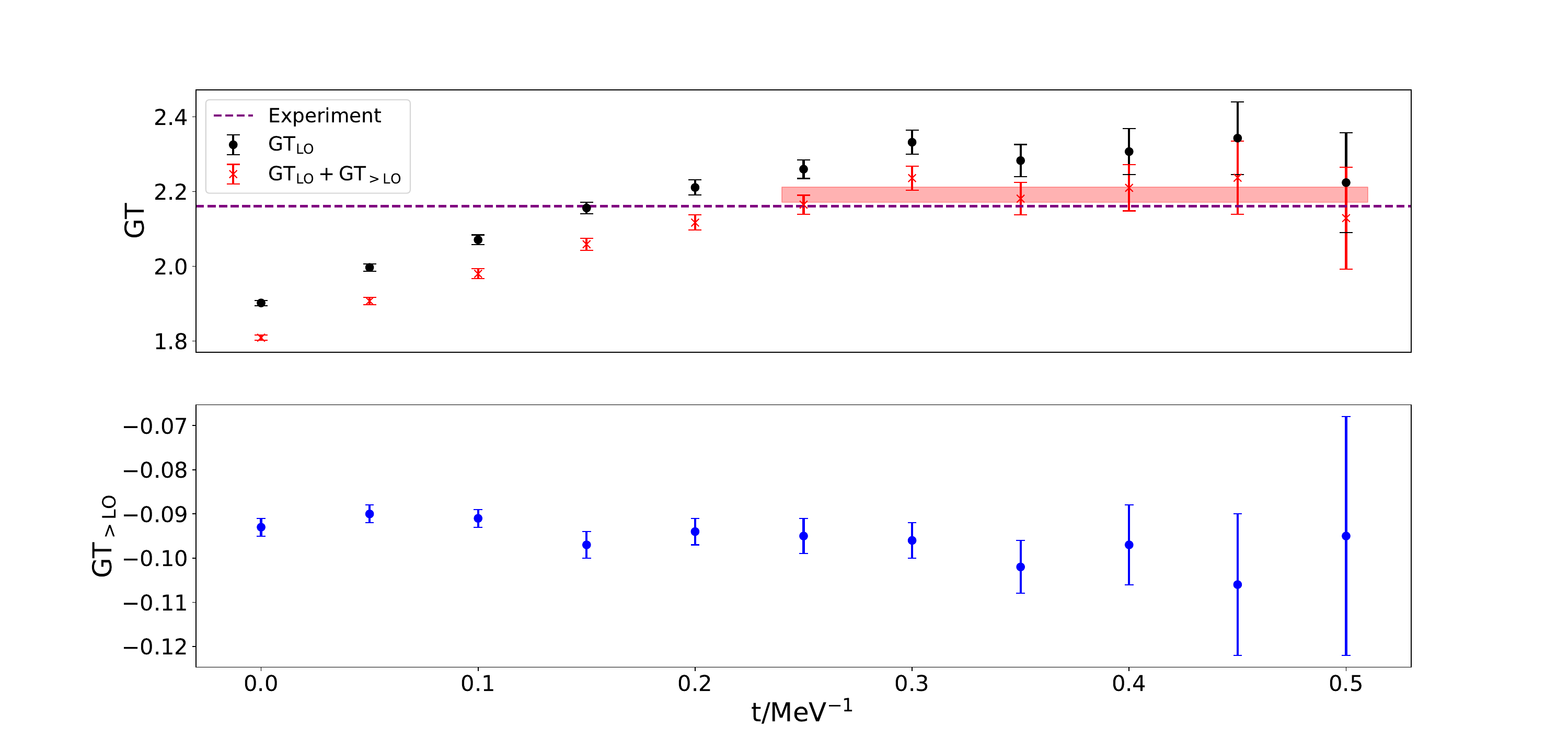}
\caption{RGTME  of $^6$He $\beta$-decay versus the Euclidean evolution time $t$. In the upper panel, the black points represent the contribution of the LO axial current, the  red points denote the full RGTME $\mathrm{GT}$ and the red band is the fitting result. The experimental value of $\mathrm{GT}$ is plotted as the dashed purple line. The lower panel shows the  net contributions from currents beyond LO. }
\label{fig:He6GT}
\end{figure*}

  \begin{table*}
  \renewcommand\arraystretch{1.2}
	\resizebox{450pt}{!}{
		\begin{tabular}{l c c c c c c c}
			\hline\hline
			    &  LO& N$^2$LO& N$^3$LO(OPE)& N$^3$LO(CT)& Total-LO& Total& Expt.~\cite{Knecht:2012}\\
			\hline
			RGTME &   2.289(19)&-0.024(2)&-0.104(1)&0.032(1)&-0.096(2)&2.192(20)&2.161(4)\\
				\hline\hline
		\end{tabular}
		}
		\caption{Contributions to RGTME of $^6$He $\beta$ decay from currents at different orders. Columns labeled with LO, N$^2$LO, N$^3$LO(OPE), and N$^3$LO(CT) correspond to the diagrams
       illustrated in panels (a), (b), (c), and (d) of Fig.~\ref{fig:current}, respectively. The full result is given
      in the column labeled ``Total," while results including only corrections beyond LO are listed under ``Total-LO." The value in parenthese is the statistical uncertainty. The experimental value in the last column is from Ref.~\cite{Knecht:2012}. }
		\label{tab:table2}
\end{table*}

To evaluate the contributions from  currents at different orders , we calculate the corresponding RGTMEs and present the results in Table~\ref{tab:table1}. It is evident that the LO current provides the dominant contribution, while corrections from higher-order currents remain below 5\%.
To validate the two computational approaches described in Sec.~\ref{sec:cal method}, we also performed Monte Carlo calculations for $^3$H $\beta$ decay. The results of both approaches, shown in the second and third rows of Table~\ref{tab:table1}, exhibit good agreement.

One key observation naturally emerges from Table~\ref{tab:table1}: our calculation yields GT$_{\mathrm{LO}}$/GT$_{\mathrm{exp}}>$1, consistent with the recent NLEFT calculation of $^3$H $\beta$ decay~\cite{Elhatisari:2024gg}. However, previous continuum-based studies~\cite{Baroni:2017, Klos:2017, Baroni:2018, Gysbers:2019} reported a value smaller than 1. The origin of this discrepancy remains unclear and warrants further investigation.

\subsection{$^6$He $\beta$ decay}

We first present the energies and charge radii of $^6$He and $^6$Li in Table~\ref{tab:He6Li6observable}. The energy is computed up to second-order perturbation theory~\cite{Lu:2021}, leading to relatively large uncertainties. The charge radius, on the other hand, is calculated up to the first order. Overall, our results show satisfactory agreement with experimental data.

Next, we present the results for $^6$He $\beta$ decay. The dependence of the RGTME on the Euclidean evolution time $t$ is illustrated in Fig.~\ref{fig:He6GT}, where  contributions from currents at LO (black points) and beyond LO (blue points) are plotted separately in the upper and lower panels, respectively. 
The full result is represented by the red points in the upper panel. For the LO current, we observe a monotonic increase with $t$ and find a plateau beginning at around $t = 0.3$ MeV$^{-1}$. In contrast, the net contribution beyond LO exhibits a relatively weaker dependence on $t$. Therefore, we make a constant fit in the interval 0.3 $\le t \le$  0.5 MeV$^{-1}$ for the LO results and 0.15 $\le t \le$  0.5 MeV$^{-1}$ for the beyond-LO results. The fit result for the full RGTME is plotted as the red band in the upper panel, and is very close to its experimental value represented by the horizontal dashed purple line.

 Table~\ref{tab:table2} presents the lattice predictions for the RGTMEs of $^6$He $\beta$-decay, including the total and partial contributions from different orders of currents. Similarly to $^3$H $\beta$ decay, the LO term remains dominant, while higher-order corrections contribute less than 5\%. When compared with experiment, the LO result is slightly larger, and the negative correction from higher-order currents shifts the  RGTME towards the correct direction and improves its agreement with the experimental data. 

Since the contributions of higher-order currents are only included with the zeroth-order perturbation theory, as presented in  Eq.~(\ref{eq:PTN2LO}), it is necessary to assess the impact of this truncation of the perturbative series. 
Our calculations indicate that the first-order perturbative correction to the LO current remains below 10\% of the zeroth-order contribution. Assuming this is also the case for higher-order currents, the truncation would affect the total RGTME by about 1\%, which is under good control and does not influence our main conclusions.

\section{Summary}
\label{sec:summary}

In this work, we investigate nuclear $\beta$ decays using lattice quantum Monte Carlo simulations within the framework of the nuclear lattice effective field theory (NLEFT). The three-body low-energy constants (LECs)
$c_D$ and $c_E$ are determined from the binding energy and the half-life of $^3$H. Our prediction for $^6$He $\beta$ decay shows reasonable agreement with experimental data and validates the method employed in this work.  

To conclude, this work establishes a foundation for investigating $\beta$ decays in light and medium-mass nuclei with $A\geq 4$ using NLEFT. We anticipate that the recently developed N$^3$LO interaction~\cite{PhysRevC.98.044002, Elhatisari:2022,arXiv2502.13565} will improve the predictions. Additionally, our approach opens the door to studying other nuclear electroweak processes of physical interest~\cite{0vbbreview1,0vbbEFT1,0vbbEFT2,Hardy:2020,superallowed2,superallowed3,superallowed4} within a unified lattice field theory framework, 
complementing lattice QCD studies at the quark and gluon level~\cite{0vbbLQCD1,0vbbLQCD2,0vbbLQCD3,gammaW2020,gammaW2023}.

\begin{acknowledgments}
We are grateful for helpful discussions on the nuclear interactions with Dean Lee, Ulf-G. Mei\ss ner, and Shihang Shen.  We appreciate Hermann Krebs for the choice of axial currents, Young-Ho Song for the design of Monte Carlo sampling method, and Jiangming Yao for helpful comments. X.F. and T.W. were supported in part by NSFC of China under Grants No. 12125501, No. 12293060, No. 12293063, and No. 12141501, and by the National Key Research and Development Program of China under Grant No. 2020YFA0406400.
B.N.L. was supported by NSAF Grant No.U2330401 and by the National Natural Science Foundation of China under Grant No.12275259.
\end{acknowledgments}

%\bibliography{beta}
\newpage

\begin{widetext}
\appendix
\setcounter{page}{1}
\renewcommand{\thepage}{Supplementary Information -- S\arabic{page}}
\setcounter{table}{0}
\renewcommand{\thetable}{S\,\Roman{table}}
\setcounter{figure}{0}
\renewcommand{\thefigure}{S\,\arabic{figure}}

\section{Nucleon-nucleon Scattering}
\label{app:a1}
We use the spherical wall method to calculate nucleon-nucleon phase shifts on the lattice~\cite{SuppNN1,SuppNN2,SuppNN3}. The fitting results of $n$-$p$ scattering for the full Hamiltonian $H$ (blue dash-dot line) and the nonperturbative Hamiltonian $H_{0}$ (black dashed line) is shown in Fig.~\ref{fig:phaseshift}, with the center-of-mass momentum $P\le 200$~MeV. We also show the phase shifts extracted from
the Nijmegen partial wave analysis~\cite{SuppNN4}, represented by the red points. For $H_{0}$, results are only fitted to $^1S_0$  and $^3S_1$ channels, so reasonable agreement with the experimental data appears in these two channels. The results for other channels are predictions based on the $S$-waves fit, and the agreement is not very satisfactory. We could in principle make an improvement by increasing the value $\Lambda'_\pi$  in Eq.~(\ref{eq:LOH}) or adding more operators in $H_{0}$, but this would introduce additional sign problems. For $H$, the fitting shows good agreement with the experiment in most channels. In several $D$-wave channels there are some deviations, which can be eliminated if  operators beyond N$^2$LO are included.  

\begin{figure}[hbtp]
\includegraphics[height=4.4in]{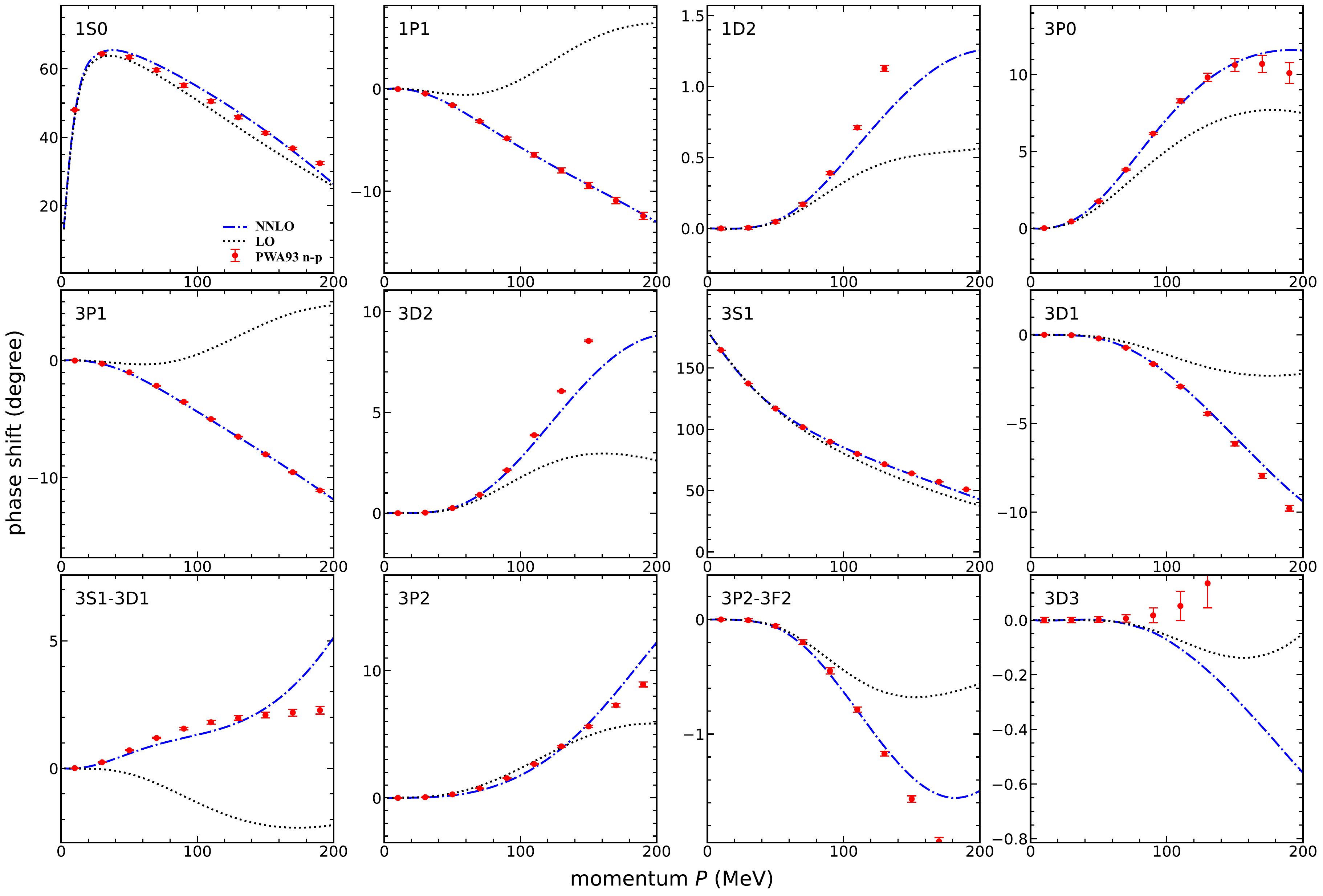}
\caption{ $n$-$p$ scattering phase shifts versus the
center-of-mass  momentum $P$ for $H$ and $H_{0}$. The red points are  from Nijmegen partial wave analysis. The blue and black lines denote fitting results of the full Hamiltonian $H$ and the nonperturbative Hamiltonian $H_{0}$, respectively. }
\label{fig:phaseshift}
\end{figure}

\begin{figure}[htbp]
\includegraphics[height=2.8in]{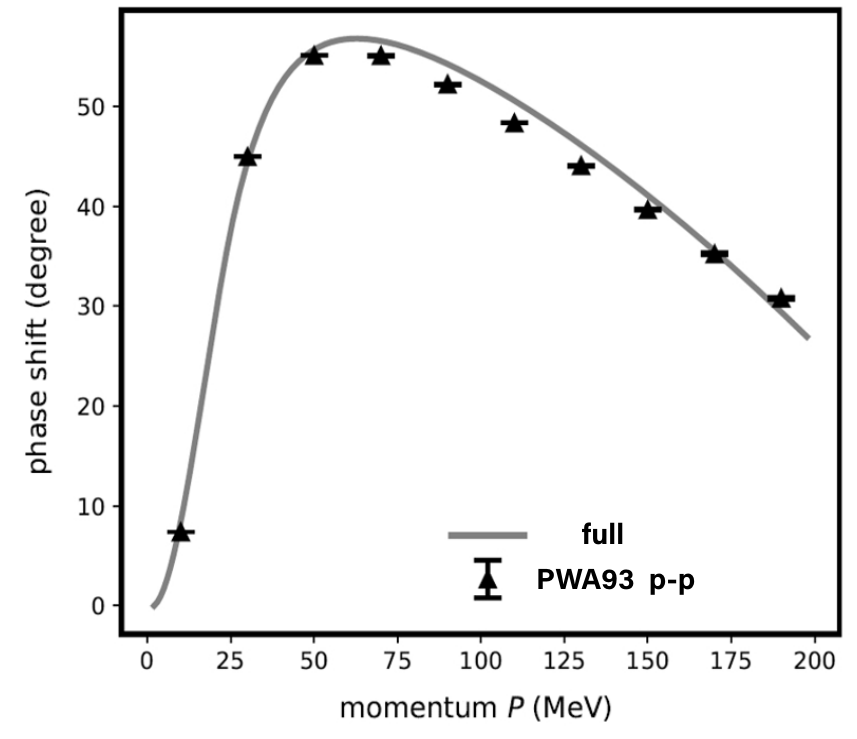}
\caption{$p$-$p$ scattering phase shifts versus the
center-of-mass  momentum $P$ in the $^1S_0$ channel. The black points with errors  are from the Nijmegen partial wave analysis. The gray line denotes the fit result.}
\label{fig:pp_phaseshift}
\end{figure}

For the isospin-breaking potentials $V_{pp}$ and $V_{nn}$, we tune the value of $C_{pp}$ to reproduce the $p$-$p$ phase shifts in the $^1S_0$ channel~\cite{SuppNN4} . The  fitting result is given  in Fig.~\ref{fig:pp_phaseshift}. For $V_{nn}$, we fit it to the experimental value of  the $n$-$n$ scattering length~\cite{SuppNN6} and effective range~\cite{SuppNN7} in the $^1 S_0$ channel:
\begin{equation}
    a_{nn}=-18.95\pm 0.40\  \mathrm{fm},\quad  r_{nn}=2.75\pm 0.11\ \mathrm{fm}.
\end{equation}

\section{Perturbative Expression of the RGTME }
\label{app:a3}
In this section, we give detailed expressions for the perturbative expansion of  the RGTME.

We write Eq.~(\ref{eq:GTlatdef}) more explicitly as
\begin{equation}
    \mathrm{GT}(t) =\frac{\sqrt{2J_f+1}}{g_A\langle J_iM_i, 11|J_fM_f\rangle}\frac{\langle \Phi_f|M^{N_t/2}\mathcal{J}M^{N_t/2}|\Phi_i\rangle}{\sqrt{\langle \Phi_i|M^{N_t}|\Phi_i\rangle}\sqrt{\langle \Phi_f|M^{N_t}|\Phi_f\rangle}}
    \label{eq:sup1}.
\end{equation}
According to Eq.~(\ref{eq: split}), the full transfer matrix is separated into two parts,
 \begin{equation}
    M = M_{0}+\Delta M.
\end{equation}
Then Eq.~(\ref{eq:sup1}) can be expanded as a power series of $\Delta M$:
\begin{equation}   \mathrm{GT}(t)=\mathrm{GT}^{(0)}(t)+\mathrm{GT}^{(1)}(t)+\mathcal{O}(\Delta M^2).
\end{equation}

To write down the expressions for $\mathrm{GT}^{(0)}$ and $\mathrm{GT}^{(1)}$, we introduce the notation
\begin{equation}
    |\Phi_{i/f, 0}^{t/2}\rangle = M_0^{N_t/2}|\Phi_{i/f}\rangle
    \label{eq:sup4}
\end{equation}
to describe the evolution of trial states under the nonperturbative Hamiltonian $H_{0}$. We also introduce
\begin{equation}
    |\delta \Phi_{i/f}^{t/2}\rangle = \sum_{k=1}^{N_t/2}M_0^{k-1}\Delta M M_0^{N_t/2-k}|\Phi_{i/f}\rangle
    \label{eq:deltaPhi}
\end{equation}
as the first-order correction to Eq.~(\ref{eq:sup4}). With Eq.~(\ref{eq:sup4}) and~(\ref{eq:deltaPhi}), it easy to derive
\begin{equation}
    \mathrm{GT}^{(0)}(t) = \frac{\sqrt{2J_f+1}}{g_A \langle J_iM_i, 11|J_fM_f\rangle}\frac{\langle \Phi_{f,0}^{t/2}|\mathcal{J}|\Phi_{i,0}^{t/2}\rangle}{\sqrt{\langle \Phi_{i,0}^{t/2}|\Phi_{i,0}^{t/2}\rangle\langle \Phi_{f,0}^{t/2}|\Phi_{f,0}^{t/2}\rangle}}
    \label{eq:GT0}
\end{equation}
and
\begin{eqnarray}
   \mathrm{GT}^{(1)}(t) &=& \frac{\sqrt{2J_f+1}}{g_A\langle J_iM_i, 11|J_fM_f\rangle}\frac{\langle \Phi_{f,0}^{t/2}|\mathcal{J}|\delta\Phi_{i}^{t/2}\rangle+\langle \delta\Phi_{f}^{t/2}|\mathcal{J}|\Phi_{i,0}^{t/2}\rangle}{\sqrt{\langle \Phi_{i,0}^{t/2}|\Phi_{i,0}^{t/2}\rangle\langle \Phi_{f,0}^{t/2}|\Phi_{f,0}^{t/2}\rangle}}\nonumber\\
    &-&\mathrm{GT}^{(0)}(t)\left[\frac{\langle \Phi_{i,0}^{t/2}|\delta\Phi_{i}^{t/2}\rangle+\langle \delta\Phi_{i}^{t/2}|\Phi_{i,0}^{t/2}\rangle}{2\langle \Phi_{i,0}^{t/2}|\Phi_{i,0}^{t/2}\rangle}+i\leftrightarrow f\right].
    \label{eq:GT1} 
\end{eqnarray}

\section{More about the Monte Carlo Method}
\label{app:a4}

In this work, we need to calculate matrix elements of the following form
\begin{equation}
    \langle \Psi^{t/2}_{a,0}|\mathcal{O}|\Psi^{t/2}_{b,0}\rangle,\ a,b\in\{i,f\}, \  \mathcal{O}\in\{1,\mathcal{J}\},
    \label{ME}
\end{equation}
as is shown in Eqs.~(\ref{eq:GT0}) and (\ref{eq:GT1}). Applying auxiliary field transformation, we have,
\begin{equation}
    \langle \Phi_{a,0}^{t/2}|\mathcal{O}|\Phi_{b,0}^{t/2}\rangle = \int\mathcal{D}s \ e^{-S[s]}Z_{a,b}^{\mathcal{O}}[s].
\end{equation}
Here $s$ is the auxiliary field located at all the lattice points and $S[s]$ is a Gaussian-type function. The interaction between the nucleons and the auxiliary field is included in $Z_{a,b}^{O}[s]$. After the transformation, both the nominators and the denominators in Eqs.~(\ref{eq:GT0}) and (\ref{eq:GT1}) become integrals over $s$. We evaluate them through generating a set of auxiliary fields which distribute according to a given probability distribution function $P[s]$. A proper choice of $P[s]$ can help improve the sampling efficiency and reduce statistical noises. As both the initial and  final states of the nucleus nucleus contribute to the $\beta$-decay process, we take the following form of $P[s]$:
\begin{equation}
    P[s]\propto c_ie^{-S[s]}|Z^{\mathcal{O}=1}_{i,i}[s]|+c_fe^{-S[s]}|Z^{\mathcal{O}=1}_{f,f}[s]|,
  \label{eq:Ps}
\end{equation}
with $c_i$ and $c_f$ positive numbers. Their values are tuned to suppress statistical errors. The update of the auxiliary field is accelerated by the shuttle algorithm proposed in Ref.~\cite{Supp2}.

\section*{}

\end{widetext}

\newpage

\end{document}